\documentclass[a4paper]{jpconf}
\usepackage{graphicx}

\usepackage{graphicx} % figures
\usepackage{psfrag} % to write in TeX on figures
\usepackage{subfigure} % multiple figures
\usepackage{array} % advanced tables
\usepackage{booktabs} % enhance quality of tables
\usepackage{amsfonts} % AMS
\usepackage{amssymb} % AMS
\usepackage{amsmath} % AMS
\usepackage{latexsym} % AMS
\usepackage{amsthm} % AMS
\usepackage{slashed} % Dirac notation
\usepackage{bm} % bold for math
\usepackage{mathrsfs} % "Script" font
\usepackage[usenames,dvipsnames]{color} % the 68 standard colors
\usepackage{subfigure}

\newcommand{\be}{\begin{equation}}
\newcommand{\ee}{\end{equation}}

\begin{document}

\title{Phenomenological tests of the Two-Higgs-Doublet Model with MFV and flavour-blind phases}

\author{Maria Valentina Carlucci}

\address{Physik-Department, Technische Universit\"at M\"unchen, \\ James-Franck-Stra{\ss}e, D-85748 Garching, Germany}

\ead{maria.carlucci@ph.tum.de}

\begin{abstract}
In the context of a Two-Higgs-Doublet Model in which Minimal Flavour Violation (MFV) is imposed, one can allow the presence of flavour-blind CP-violating phases without obtaining electric dipole moments that overcome the experimental bounds. This choice permits to accommodate the hinted large phase in the $B_s$ mixing and, at the same time, to soften the observed anomaly in the relation between $\epsilon_K$ and $S_{\psi K_S}$.
\end{abstract}

%%%%%%%%%%%%%%%%%%%%%%%%%%%%%%%%%%
\section{Introduction}
%%%%%%%%%%%%%%%%%%%%%%%%%%%%%%%%%%

During the last decade the $B$ factories and the Tevatron experiments have collected a large amount of data that permits a deep analysis of precision measurements. The agreement of the experimental data of flavour-changing processes with the Standard Model (SM) predictions is verified within 20\%; a similar agreement is found with the CP-violating processes, but recently a few anomalies have been detected in this sector: (i) the CP-violation in the $B_s$ system signaled by the CP-asymmetry $S_{\psi \phi}$ in $B_s \rightarrow \psi \phi$ observed by CDF and D0 that appears to be roughly by a factor of 20 larger than the SM and MFV predictions \cite{Aaltonen:2007he,Abazov:2008fj}; (ii) the value of $\sin 2\beta$ resulting from the UT fits tends to be significantly larger than the measured value of $S_{\psi K_S}$ \cite{Buras:2008nn}; (iii) the value of $\epsilon_K$ predicted in the SM by using $S_{\psi K_S}$ as the measure of the observed CP-violation is about $2\sigma$ lower than the data \cite{Buras:2008nn,Lunghi:2008aa}; (iv) recently D0 reported a measurement of the like-sign dimuon charge asymmetry in semileptonic $b$ decay that is $3.2\sigma$ from the SM prediction \cite{Abazov:2010hv}. These anomalies, if confirmed, suggest that a room for New Physics could be present in CP-violating processes, even if other CP-violating observables, such as electric dipole moments (EDMs), put tight constraints on it.

One of the most elegant and effective ways to protect the flavour in the New Physics models is the MFV hypothesis \cite{Buras:2000dm,D'Ambrosio:2002ex}; it forces the flavour sector of the model to respect the same flavour symmetry and symmetry breaking as the SM, i.e.\ to be governed only by the same Yukawa couplings. However, while in the SM the Yukawa couplings are the only sources of both flavour breaking and CP violation, this is not necessary true in New Physics models even assuming MFV: the flavour can be broken only by Yukawa couplings, but other sources of CP violation are allowed \cite{Kagan:2009bn}. This possibility is particularly interesting in light of the experimental results described above, in which the flavour structure of the SM seems to be confirmed, but new CP-violating mechanisms could contribute.

We have analyzed a simple extension of the SM in which two Higgs doublets are present; we have imposed MFV but allowed the presence of flavour-blind CP-violating phases in the fermion-scalar interactions; we will call this model 2HDM$_{\overline{\mathrm{MFV}}}$ \cite{Buras:2010mh}. First of all we have checked that the strict experimental flavour-changing neutral current (FCNCs) constraints can be naturally fulfilled, and that the bounds on CP-violation that come from EDMs are not overcome \cite{Buras:2010zm}. Then we have shown that a large phase in the $B_s$ mixing can be easily accommodated and this, without extra free parameters, improves significantly in a correlated manner the issues about $\epsilon_K$ and $S_{\psi K_S}$.

%%%%%%%%%%%%%%%%%%%%%%%%%%%%%%%%%%
\section{The 2HDM$_{\overline{\mathrm{MFV}}}$}
%%%%%%%%%%%%%%%%%%%%%%%%%%%%%%%%%%

%%%%%%%%%%%%%%%%%%%%%%%%%%%%%%%%%%
\subsection{General structure of 2HDMs}
%%%%%%%%%%%%%%%%%%%%%%%%%%%%%%%%%%

The Higgs Lagrangian of a generic  model with two-Higgs doublets, $H_1$ and 
$H_2$, with hypercharges $Y=1/2$ and $Y=-1/2$ respectively, can be written as
\be
\mathcal{L} = \sum_{i=1,2} D_\mu H_i D^\mu H_i^\dagger + 
\mathcal{L}_{Y}  - V(H_1,H_2)~,
\ee
where $D_\mu H_i = \partial_{\mu}H_i-i g^\prime Y \hat{B}_{\mu}H_i -ig T_a\hat{W^a}_{\mu} H_i$, with $T_a=\tau_a/2$. 

The potential $V(H_1,H_2)$ is such that the $H_i$ gets vacuum expectation value $\langle H^0_{1(2)} \rangle = v_{1(2)}$ with $v = \sqrt{ v_1^2 + v_2^2 } \approx 246 \text{ GeV}$ fixed by the mass of the $W$ boson; moreover, we only consider the case in which it does not contain new sources of CP violation. The Higgs spectrum contains three Goldstone bosons $G^{\pm}$ and $G^0$, two charged Higges $H^{\pm}$, and three neutral Higgses $h^0$, $H^0$ (CP-even), and $A^0$ (CP-odd).

The most general renormalizable and gauge-invariant interaction of the two Higgs doublets with the SM quarks is 
\be
- \mathcal{L}_Y = \bar Q_L X_{d1} D_R H_1 + \bar Q_L X_{u1} U_R H_1^c 
+ \bar Q_L X_{d2} D_R H_2^c + \bar Q_L X_{u2} U_R H_2 +{\rm h.c.}~,
\ee
where $H_{1(2)}^c = -i\tau_2 H_{1(2)}^*$  and the $X_i$ are $3\times 3$ matrices with a generic flavour structure. By performing a global rotation of angle $\beta = \text{arctan} (v_2/v_1)$ of the Higgs fields $(H_1, H_2)$ to the so-called Higgs basis $(\Phi_v, \Phi_H)$, the mass terms and the interaction terms are separated:
\be
- \mathcal{L}_Y = \bar Q_L \left( \frac{\sqrt{2}}{v} M_{d} \Phi_v + Z_{d} \Phi_H \right)D_R + \bar Q_L \left(\frac{\sqrt{2}}{v} M_{u} \Phi_v^c + Z_{u} \Phi_H^c \right) U_R  +{\rm h.c.}~;
\ee
the quark mass matrices $M_{u,d}$ and the couplings $Z_{u,d}$ are linear combinations of the $X_i$, weighted by the Higgs vacuum expectation values:
\be
M_{u,d} = \frac{v}{\sqrt{2}} \left( \cos \beta X_{{u,d}\,1} + \sin \beta X_{{u,d}\,2} \right)~, \qquad Z_{u,d} = \cos \beta X_{{u,d}\,2} - \sin \beta X_{{u,d}\,1}~.
\ee
In this way it is clear that $M_{u,d}$ and $Z_{u,d}$ cannot be diagonalized simultaneously for generic $X_i$, and we are left with dangerous FCNC couplings to the neutral Higgses.

%%%%%%%%%%%%%%%%%%%%%%%%%%%%%%%%%%
\subsection{Minimal Flavour Violation}
%%%%%%%%%%%%%%%%%%%%%%%%%%%%%%%%%%

In order to suppress the FCNCs, some additional hypotheses on the model are needed. The oldest but still widely used method to obtain the aim is the Natural Flavour Conservation hypothesis \cite{Glashow:1976nt}, that states that renormalizable couplings contributing at the tree-level to FCNC processes are simply absent by assumption. Hence, in order to forbid these terms, one should impose some additional symmetry; however, it has been shown in \cite{Buras:2010mh} that this structure is not stable under quantum corrections, and sizable FCNCs operators are generated at higher energies unless a very severe fine tuning is performed.

It seems instead that a protection of the flavour symmetry and of its breaking is necessary in many New Physics models. In this frame the popular hypothesis of MFV turns out to be effective in suppressing FCNCs and stable under renormalization. Formally, MFV consists in the assumption that the $SU(3)$ quark flavour symmetry is broken only by two independent terms, $Y_d$ and $Y_u$, transforming as
\be
Y_u \sim (3, \bar 3,1)_{SU(3)^3}~, \qquad Y_d \sim (3, 1, \bar 3)_{SU(3)^3}~;
\ee
in the 2HDM it implies for the $X_i$ the structure \cite{D'Ambrosio:2002ex}
\begin{subequations}
\begin{align}
X_{d1} &= Y_d \\
X_{d2} &= P_{d2}(Y_u Y_u^\dagger, Y_d Y_d^\dagger) \times Y_d = \epsilon_{0} Y_d + \epsilon_{1} Y_d  Y_d^\dagger Y_d                    
+  \epsilon_{2}  Y_u Y_u^\dagger Y_d + \ldots \\
X_{u1} &= P_{u1}(Y_u Y_u^\dagger, Y_d Y_d^\dagger) \times Y_u = \epsilon^\prime_{0} Y_u + \epsilon^\prime_{1}  Y_u Y_u^\dagger Y_u 
+  \epsilon^\prime_{2}  Y_d Y_d^\dagger Y_u + \ldots \\
X_{u2} &= Y_u
\end{align}
\end{subequations}
that is renormalization group invariant. Differently from the Natural Flavour Conservation case, now FCNCs are absent only at the lowest order in $Y_i Y_i^\dagger$ since the $X_i$ are aligned \cite{Pich:2009sp}; they are present when one considers higher orders instead.

In order to investigate these FCNCs, one can perform an expansion in powers of suppressed off-diagonal CKM elements, so that the effective non-diagonal $Z_d$ between the down-type quarks and the neutral Higgses assumes the form \cite{D'Ambrosio:2002ex}
\be
Z_d = W^d \lambda_d \qquad \mathrm{with} \qquad W^{d}_{ij} = \bar{a} \, \delta_{ij} + \left( a_0 V^\dagger \lambda_u^2 V + a_1 V^\dagger \lambda_u^2 V  \Delta + a_2  \Delta V^\dagger \lambda_u^2 V \right)_{ij} ~,
\ee
where $\lambda_{u,d} \propto 1/v \; \text{diag} \left( m_{u,d},m_{c,s},m_{t,b} \right)$, $\Delta = \text{diag} \left( 0,0,1 \right)$, and the $\bar{a}$, $a_i$ are parameters naturally of $\mathcal{O} (1)$; this structure already shows a large suppression due to the presence of two off-diagonal CKM elements and the down-type Yukawas. The case of $Z_u$ is less interesting since other suppression are present, and the FCNCs are negligible.

One can check that no fine tuning is required in this case by performing a comparison with experimental FCNCs observables: constraints on the free parameters are obtained by imposing that the new physics contributions must be compatible with the experimental data within errors. We have found the bounds \cite{Buras:2010mh}:
\begin{subequations}
\begin{align}
& |a_0| \tan\beta \frac{v}{M_H } < 18 && \text{from} \; \epsilon_K~, \\
& \sqrt{|(a^*_0+a^*_1)(a_0+a_2)|} \tan\beta \frac{v}{M_H } = 10 && \text{from} \; \Delta M_s~, \\
& \sqrt{|a_0+a_1|} \tan\beta \frac{v}{M_H } < 8.5 && \text{from} \; \text{Br}\left( B_s \rightarrow \mu^+ \mu^- \right)~;
\end{align}
\end{subequations}
that, as can be noted, are well compatible and perfectly natural.

%%%%%%%%%%%%%%%%%%%%%%%%%%%%%%%%%%
\subsection{Flavour-blind phases}
%%%%%%%%%%%%%%%%%%%%%%%%%%%%%%%%%%

The mechanisms of flavour and CP violation do not necessary need to be related: in MFV the Yukawa matrices are the only sources of flavour breaking, but other sources of CP violation could be present, provided that they are flavour-blind \cite{Kagan:2009bn}. However, so far it has been assumed the effective MFV parameter{\bf s $a_i$} to be real, in order to fulfill the strong bounds on flavour-conserving CPV phases implied by the electric dipole moments. Allowing the FCNC parameters $a_i$ to be complex, we investigate the possibility of generic CP-violating flavour-blind phases in the Higgs sector \cite{Buras:2010mh}; we will show that this choice implies several interesting phenomenological results for the $\Delta F = 2$ transitions, and that at the same time the electric dipole moments, although much enhanced over the SM values, are still compatible with the experimental bounds.

%%%%%%%%%%%%%%%%%%%%%%%%%%%%%%%%%%
\section{$\Delta F = 2$ amplitudes}
%%%%%%%%%%%%%%%%%%%%%%%%%%%%%%%%%%

Considering the $\Delta F = 2$ FCNC transitions mediated by the neutral Higgs bosons, the leading MFV effective Hamiltonians are:
\begin{subequations}
\begin{align}
\mathcal{H}^{|\Delta S|=2} & \propto
 - \frac{|a_0|^2}{  M_H^2} ~\frac{m_s}{v} \frac{m_d}{v} \left[ \left( \frac{m_t}{v} \right)^2 V^*_{ts} V_{td} \right]^2~
 ({\bar s}_R d_L) ({\bar s}_L d_R)  {\rm ~+~h.c.}~, \\
\mathcal{H}^{|\Delta B|=2}_{d,s} & \propto
 - \frac{(a^*_0+a^*_1)(a_0+a_2)}{ M_H^2} ~\frac{m_b}{v} \frac{m_{d,s}}{v} \left[ \left( \frac{m_t}{v} \right)^2 V^*_{tb} V_{t(d,s)} \right]^2~
 \left[{\bar b}_R (d,s)_L \right] \, \left[{\bar b}_L (d,s)_R \right]  {\rm ~+~h.c.}~,
\end{align}
that show two key properties:
\end{subequations}
\begin{itemize}
\item the impact in $K^0$,  $B_{d}$ and $B_{s}$ mixing amplitudes scales with $m_sm_d$, $m_b m_d$ and $m_b m_s$ respectively, opening the possibility of sizable non-standard contributions to the $B_s$ system  without serious constraints from  $K^0$ and  $B_{d}$ mixing;
\item while the possible flavour-blind phases do not contribute to the $\Delta S=2$ effective Hamiltonian, they could have an impact in the $\Delta B=2$ case, offering the possibility to solve the anomaly in the $B_s$ mixing phase.
\end{itemize}
These Hamiltonians have a direct impact on some crucial observables of the neutral mesons systems, namely the mass differences and the asymmetries in the decays; in fact, in this analysis we consider
\begin{align}
& \Delta M_K = \frac{1}{m_K} \mathrm{Re} \left[ \left\langle K^0 \right| \mathcal{H}^{|\Delta S|=2} \left| \bar{K}^0 \right\rangle \right]~, \\
& \epsilon_K \simeq e^{i \frac{\pi}{4}} \frac{1}{ \sqrt{2} \, m_K \Delta M_K} \mathrm{Im} \left[ \left\langle K^0 \right| \mathcal{H}^{|\Delta S|=2} \left| \bar{K}^0 \right\rangle \right] ~, \\
& \Delta M_{d,s} = \frac{1}{m_{B_{d,s}}} \left| \left\langle B_{d,s}^0 \right| \mathcal{H}^{|\Delta B|=2}_{d,s} \left| \bar{B}_{d,s}^0 \right\rangle \right|~, \\
& S_{\psi K_S} = \sin \left( \arg \left\langle B_d^0 \right| \mathcal{H}^{|\Delta B|=2}_{d} \left| \bar{B}_d^0 \right\rangle \right)~, \\
& S_{\psi \phi} = \sin \left( \arg \left\langle B_s^0 \right| \mathcal{H}^{|\Delta B|=2}_{s} \left| \bar{B}_s^0 \right\rangle \right)~.
\end{align}

Using the presence of the new free phase $\mathrm{arg} \left[ (a^*_0+a^*_1)(a_0+a_2) \right]$ in $\mathcal{H}^{|\Delta B|=2}_s$, the hinted large value of $S_{\psi\phi}$ can be easily obtained. Moreover, MFV implies that the new phases in the $B_d$ and the $B_s$ systems are related by the ratio $m_d/m_s$, and hence a large phase in the $B_s$ system determines an unambiguous small shift in the relation between $S_{\psi K_S}$ and the CKM phase $\beta$; as it can be seen in Fig.~1 (Left), it goes in the right direction to improve the existing tension between  the experimental value of $S_{\psi K_S}$ and its SM prediction.

\begin{figure}[h]
\centering
\subfigure{\includegraphics[width=0.45\textwidth]{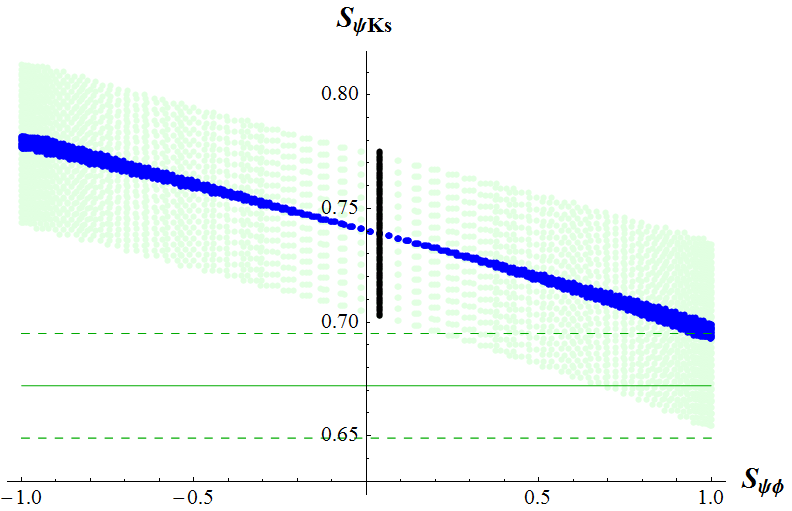}} $\qquad$
\subfigure{\includegraphics[width=0.45\textwidth]{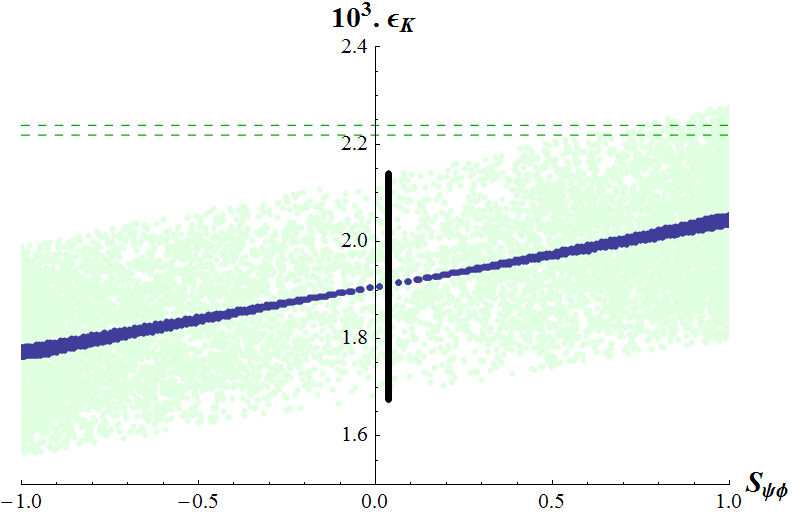}}
\caption{(Left) Correlation between $S_{\psi K_S}$ and $S_{\psi\phi}$. The dark points have been obtained with the CKM phase $\beta$ fixed to its central value: the spread is determined only by the requirement of a deviation of $\Delta M_s$ within 10\% of its SM value; the light points represent the $\pm 1\sigma$ error due to the uncertainty in the extraction of $\beta$. The $\pm1\sigma$ range of $\phi_{B_S}^{\text{exp}}$ (light horizontal lines) and the SM prediction (black vertical line) are also shown. (Right) Correlation between $\epsilon_K$ and $S_{\psi K_S}$. Notations as before.}
\end{figure}

Due to the $m_sm_d$ factor in $\mathcal{H}^{|\Delta S|=2}$, the new physics contribution to $\epsilon_K$ is tiny and does not improve alone the agreement between data and prediction for $\epsilon_K$. However, given the modified relation between $S_{\psi K_S}$ and the CKM phase $\beta$, the true value of $\beta$ extracted in this scenario increases with respect to SM fits. As a result of this modified value of $\beta$, also the predicted value for $\epsilon_K$ increases with respect to the SM case, resulting in a better agreement with data, as shown in Fig.~1 (Right).

%%%%%%%%%%%%%%%%%%%%%%%%%%%%%%%%%%
\section{Electric dipole moments}
%%%%%%%%%%%%%%%%%%%%%%%%%%%%%%%%%%

%%%%%%%%%%%%%%%%%%%%%%%%%%%%%%%%%%
\subsection{2HDM$_{\overline{\mathrm{MFV}}}$ predictions}
%%%%%%%%%%%%%%%%%%%%%%%%%%%%%%%%%%

The EDM of a particle is defined by one of its electromagnetic form factors \cite{Bernreuther:1990jx}. In particular, for a spin-$\frac{1}{2}$ particle $f$, the form-factor decomposition of the matrix element of the electromagnetic current $J_{\mu}$ is
\be
\left\langle f(p') \right| J_{\mu} (0) \left| f(p) \right\rangle = \bar{u} (p') \Gamma_{\mu} (p'-p) u (p)
\ee
where
\be
\Gamma_{\mu} (q) = F_1 (q^2) \gamma_{\mu} + F_2 (q^2) i \sigma_{\mu \nu} \frac{q^{\nu}}{2m} + F_A (q^2) \left( \gamma_{\mu} \gamma_5 q^2 - 2m \gamma_5 q_{\mu} \right) + F_3 (q^2) \sigma_{\mu \nu} \frac{q^{\nu}}{2m} ~.
\ee
Since the electric dipole interaction of a particle with the electromagnetic field violates both $P$ and $T$ symmetry, one identifies the electric dipole moment of $f$ with
\be
d_f \equiv - \frac{F_3 (0)}{2m} ~;
\ee
this corresponds to the effective electric dipole interaction
\be
\mathcal{L}_{\mathrm{eff}} = - \frac{i}{2} d_f \bar{\psi} \sigma_{\mu \nu} \gamma_5 \psi F^{\mu \nu} ~,
\ee
that indeed reduces to $\mathcal{L}_{\mathrm{eff}} = d_f \, \vec{\sigma} \cdot \vec{E}$ in the nonrelativistic limit. In an analogous way one can define the chromoelectric dipole moment. Moreover, beyond the one-loop single-particle level, there are other CP-violating operators that can have significant effects on the EDMs, like, for example, the dimension-6 Weinberg operators or the CP-odd four-fermion interactions.

The effective Lagrangian describing the fermions (C)EDMs that is relevant for our analysis reads
\be
-\mathcal{L}_{\mathrm{eff}} = \sum_{f} i \frac{d_f}{2}\bar{f} F_{\mu \nu} \sigma^{\mu \nu} \gamma_5 f
+ \sum_{f} i \frac{d^c_f}{2} g_s \bar{f} G_{\mu \nu} \sigma^{\mu \nu} \gamma_5 f + \sum_{f,f'}C_{ff'}(\bar{f}f)(\bar{f'}i\gamma_5f') ~,
\ee
where $d^{(c)}_{f}$ stands for the quarks and leptons (C)EDMs, while $C_{ff'}$ is the coefficient of the CP-odd four fermion interactions.

Among the various atomic and hadronic EDMs, the Thallium, neutron and Mercury ones represent the most sensitive probes of CP violating effects; their EDMs can be calculated from the elementary particles EDMs using non-perturbative techniques \cite{Demir:2003js}. The thallium EDM ($d_{\rm{Tl}}$) can be estimated as
\be
d_{\mathrm{Tl}} \simeq -585\cdot d_e - e\,\left(43\rm{GeV}\right)\,C_{S} ~,
\ee
where
\be
C_S \simeq C_{de}\frac{29\,{\rm MeV}}{m_d} + C_{se} \frac{\kappa \times 220\,{\rm MeV}}{m_s} + C_{be}\frac{ 66\,{\rm MeV}(1-0.25 \kappa)}{m_b} ~,
\ee
with $\kappa \simeq 0.5 \pm 0.25$; the neutron and mercury EDMs can be estimated from QCD sum rules, which leads to
\be
 d_n = (1\pm 0.5)\Big[ 1.4 (d_d-0.25 d_u) + 1.1 e\, (d^c_d+0.5 d^c_u) \Big]
\ee
and
\begin{multline}
d_{\mathrm{Hg}} \simeq 7\times 10^{-3}\,e\,(d_u^c-d_d^c) + 10^{-2}\,d_e + e\,\left( 3.5 \times 10^{-3}\,\rm{GeV} \right)\,C_{S} - \\
- e\left( 1.4 \times 10^{-5}\rm{GeV}^2 \right) \left[0.5\frac{C_{dd}}{m_d} + 3.3 \kappa \frac{C_{sd}}{m_s} + (1-0.25 \kappa)\frac{C_{bd}}{m_b}\right] 
\end{multline}
(in which the $d^{(c)}_{f}$ are evaluated at $1$~GeV).

The values of $C_{ff'}$ and $d^{(c)}_f$ have been obtained in the context of Supersymmetry in \cite{Ellis:2008zy,Hisano:2008hn}, and they have been extended to the case of the 2HDM$_{\overline{\mathrm{MFV}}}$ in \cite{Buras:2010zm}; here we present the results for the case in which CP is not violated in the scalar potential.

\begin{itemize}
\item For $C_{ff'}$ one has
\be
C_{ff'} = \frac{m_{f}\,m_{f'}}{v^2}\, \frac{\mathrm{Im}\,\omega_{ff'}}{M^2_A}\,t^{2}_{\beta}\,,
\ee
with
\be
\omega_{ff'} \simeq W^{d\star}_{ff}\,W^{d}_{f'f'} ~.
\ee
The explicit expressions for the $\omega_{ff'}$ relevant for our analysis are
\begin{align}
\mathrm{Im} \,\omega_{ed} &= {\rm Im}\,\omega_{es} = \mathrm{Im} \,\sigma_d ~, \\
\mathrm{Im} \,\omega_{eb}&= \mathrm{Im}\,\xi ~, \\
\mathrm{Im}\,\omega_{dd}&= \mathrm{Im}\,\omega_{ds}= 0 ~, \\
\mathrm{Im}\,\omega_{db}&= \mathrm{Im}\,(\sigma^{\star}_d \, \xi) ~,
\end{align}
where we have defined
\be
\xi = \overline{a} + (a_0 + a_1 + a_2)\lambda_t^2 \,, \qquad \quad
\sigma_q = \overline{a} + |V_{tq}|^2 \lambda_t^2 a_0 ~.
\ee
\item Concerning the fermions (C)EDMs, for the down quark they are induced already at the one-loop level by means of the exchange of the charged-Higgs boson and top quark. However, these effects are CKM suppressed by the factor $|V_{td}|^2\approx 10^{-4}$, and even the most extreme choice of the parameters leads to predictions for $d_n$ and $d_{\rm Hg}$ well under control \cite{Buras:2010zm}; hence, we can neglect these contributions in this analysis.

Instead, the two-loop Barr--Zee contributions \cite{Barr:1990vd} to the fermionic (C)EDMs dominate over one loop-effects since they overcome the strong CKM suppression; in the 2HDM$_{\overline{\mathrm{MFV}}}$ they read
\begin{align}
\frac{d_f}{e} &= -\sum_{q=t,b} \frac{N_c q_f \alpha_{\rm em}^2\,q_q^2\,m_f}{8\pi^2s_W^2M_W^2} \frac{m_q^2}{M_{A}^2}\,(\tan \beta)^{0,2} \left[ f(\tau_{q})\,{\rm Im}\,\omega_{qf} + g(\tau_{q})\,{\rm Im}\,\omega_{fq} \right]\,, \\
d_{f}^c &= -\sum_{q=t,b} \frac{\alpha_s\,\alpha_{\rm em}\,m_{f}}{16\pi^2s_W^2M_W^2} \frac{m_q^2}{M_{A}^2}\,(\tan \beta)^{0,2} \left[ f(\tau_{q})\,{\rm Im}\,\omega_{qf} + g(\tau_{q})\,{\rm Im}\,\omega_{fq} \right] \,,
\end{align}
where $q_{\ell}$ is the electric charge of the fermion $\ell$, $\tau_{q}=m_q^2/M_{A}^2$ and $f(\tau)$, $g(\tau)$ are the two-loop Barr--Zee functions
\begin{align}
f(\tau) & = \frac{\tau}{2} \int_0^1 \frac{1-2x(1-x)}{\tau-x(1-x)} \log \left[ \frac{x(1-x)}{\tau} \right] ~, \\
g(\tau) & = \frac{\tau}{2} \int_0^1 \frac{1}{\tau-x(1-x)} \log \left[ \frac{x(1-x)}{\tau} \right] ~,
\end{align}
and the relevant $\omega_{ff'}$ are only
\begin{align}
\mathrm{Im} \,\omega_{et}&=0 ~,\\
\mathrm{Im} \,\omega_{dt}&=\mathrm{Im}\,\sigma^{\star}_d ~,
\end{align}
since, as we have already noted, the MFV structure suppresses the contributions of the $u$ quark with a factor $( \tan \beta )^{-2}$.
\end{itemize}

A comparison of the 2HDM$_{\overline{\mathrm{MFV}}}$ predictions for the Thallium, neutron and Mercury EDMs with their present experimental upper bounds is shown in Tab.~1. They have been obtained with the reference values $m_A = 500$ GeV and $\tan \beta = 10$; it can be seen that for natural values of $|\bar{a}|,|a_i| = \mathcal{O} (1)$ they are compatible with the experimental limits.

\begin{table}[htb]
\centering
\begin{tabular}{lcl}
\toprule
Observable & Experimental bounds & 2HDM$_{\overline{\mathrm{MFV}}}$ predictions \\
\midrule
$|d_{Tl}| \quad$[$e$ cm]	& $< 9.0 \times 10^{-25}$	& $3 \times 10^{-25} \, \mathrm{Im} \,\xi + 5 \times 10^{-25} \, \mathrm{Im} \, \sigma_d$ \\
$|d_{n}| \quad$[$e$ cm]	& $< 2.9 \times 10^{-26}$	& $10^{-27} \, \mathrm{Im} \, ( \sigma_d^{\star} \, \xi)$ \\
$|d_{Hg}| \quad$[$e$ cm]	& $< 3.1 \times 10^{-29}$	& $4 \times 10^{-29} \, \mathrm{Im} \, ( \sigma_d^{\star} \, \xi) + 5 \times 10^{-29} \, \mathrm{Im} \,\sigma_d + 2 \times 10^{-29} \, \mathrm{Im} \, \xi$ \\
\bottomrule
\end{tabular}
\caption{Predictions of the 2HDM$_{\overline{\mathrm{MFV}}}$ for some experimentally interesting EDMs.}
\end{table}

%%%%%%%%%%%%%%%%%%%%%%%%%%%%%%%%%%
\subsection{Correlation between EDMs and CP violation in the $B_s$ mixing}
%%%%%%%%%%%%%%%%%%%%%%%%%%%%%%%%%%

In the previous section we have shown that, in the context of the 2HDM$_{\overline{\mathrm{MFV}}}$, one can accommodate a larger $B_s$ mixing phase obtaining at the same time some other interesting effects. However, it is necessary to check if this new phase can cause disagreements with other CP violating observables, such as the EDMs. 

\begin{figure}[htb]
\centering
\includegraphics[width=0.45\textwidth]{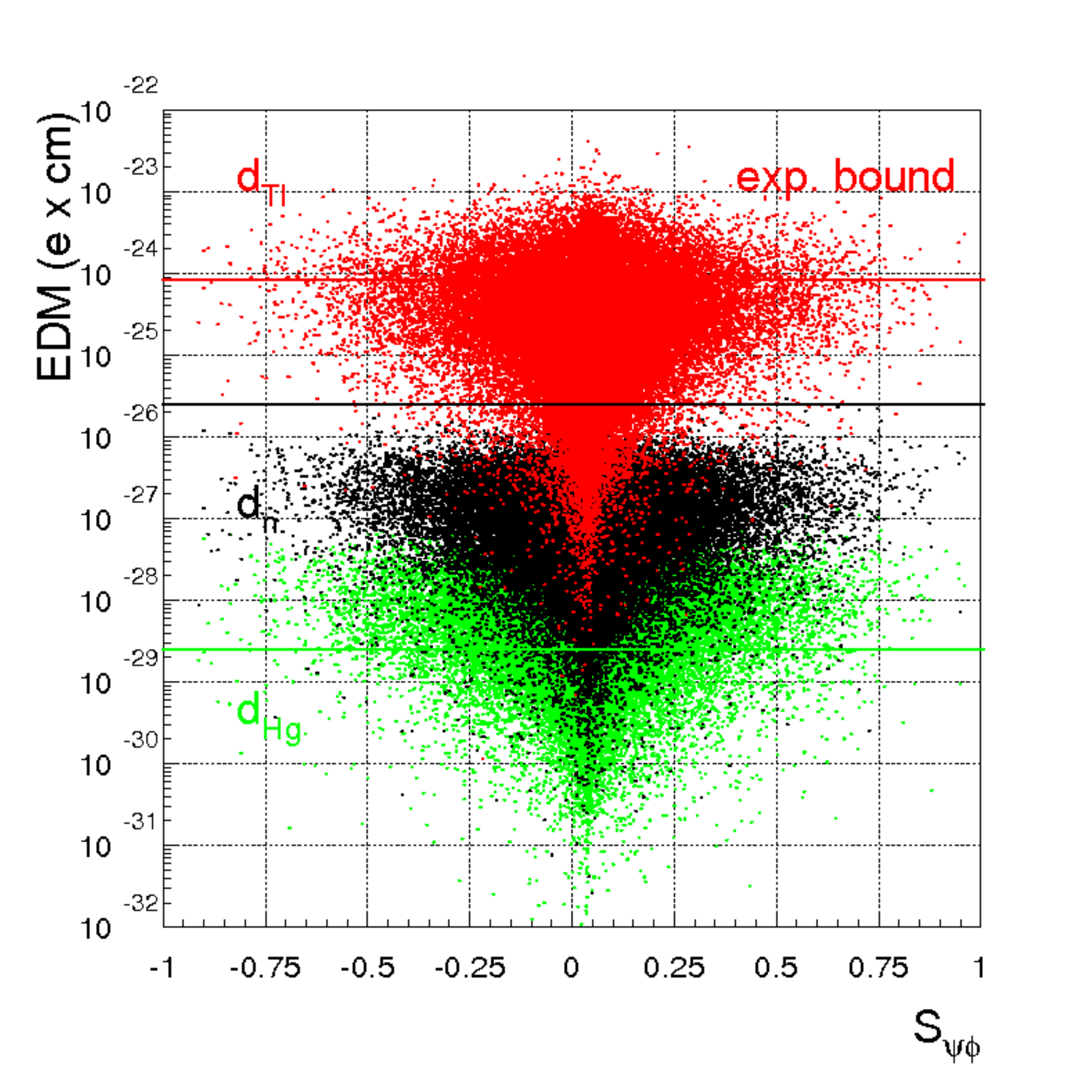}
\caption{Correlation between EDMs of the Thallium (red dots), neutron (black dots), and Mercury (green dots), versus $S_{\psi\phi}$ \cite{Buras:2010zm}. It is obtained by means of the following scan: $(|\overline{a}|,|a_0|,|a_1|,|a_2|)< 2$, $0<(\phi_{\overline{a}},\phi_{a_0},\phi_{a_1},
\phi_{a_2})< 2\pi$, $\tan\beta < 60$, $M_{H^{\pm}}< 1.5$~TeV.}
\end{figure}

The correlation plot between the considered EDMs and the observable $S_{\psi \phi}$ is shown in Fig.~2. As it can be seen, the current constraints on the EDMs still allow values of $|S_{\psi\phi}|$ larger than 0.5, compatible with the highest values of the $B_s$ mixing phase reported by the Tevatron experiments.

%%%%%%%%%%%%%%%%%%%%%%%%%%%%%%%%%%
\section{Conclusions}
%%%%%%%%%%%%%%%%%%%%%%%%%%%%%%%%%%

It is well known that the choice of introducing only one Higgs doublet in the Standard Model is just the most economical, but not the only possible one; there is a variety of New Physics models that contain more Higgs doublets, bringing interesting phenomenological features such as new sources of CP violation, dark matter candidates, axion phenomenology. In order to protect 2HDMs from FCNCs the application of MFV is natural and effective, and still allows the introduction of new flavour-blind CP-violating phases whose presence is interesting in order to address the recent experimental anomalies detected in this sector. In the 2HDM$_{\overline{\mathrm{MFV}}}$ in fact, once a larger phase in the $B_s$ mixing is introduced, the tensions in the relation between $\epsilon_K$, the CKM angle $\beta$ and the asymmetry $S_{\psi K_S}$ are automatically softened, and the agreement of the predictions of flavour-violating processes (FCNCs) and flavour-conserving observables (EDMs) with experimental data is not spoiled.

Experimental data to test the 2HDM$_{\overline{\mathrm{MFV}}}$ will probably be available in the next few years: first of all, the predicted values for the EDMs are strongly enhanced with respect to the SM, and sizable non standard values for $S_{\psi\phi}$ imply lower bounds for the aforementioned EDMs within the reach of the expected future experimental resolutions; moreover, the MFV structure of the model determines several unambiguous correlations between observables that are going to be studied by future experiments, such as the decays $B^0_{d,s} \rightarrow \mu^+ \mu^-$.

%%%%%%%%%%%%%%%%%%%%%%%%%%%%%%%%%%
\section*{Acknowledgments}
%%%%%%%%%%%%%%%%%%%%%%%%%%%%%%%%%%

I would like to thank Andrzej J. Buras, Gino Isidori and Stefania Gori for the fruitful collaboration, Paride Paradisi for valuable explanations, and Marco Bardoscia for revisions. This work has been supported in part by the Graduiertenkolleg GRK 1054 of DFG and in part by the Cluster of Excellence `Origin and Structure of the Universe'.

%%%%%%%%%%%%%%%%%%%%%%%%%%%%%%%%%%
\section*{References}
%%%%%%%%%%%%%%%%%%%%%%%%%%%%%%%%%%

\end{document}